\documentclass[runningheads]{llncs}

\usepackage[T1]{fontenc}
\usepackage{graphicx}
\usepackage{booktabs}
\usepackage{amssymb}

\usepackage[hidelinks]{hyperref}

\usepackage{color}

\begin{document}
\title{\emph{Cat}ching The \emph{C}orrect \emph{A}nswer \emph{T}rap: Characterising AI Tutor Blind Spots When Analysing Student Reasoning}

\titlerunning{Catching the Correct Answer Trap}

\author{Moiz Imran\inst{1}\and
Sahan Bulathwela\inst{1,2}}
\authorrunning{M. Imran and S. Bulathwela}
\institute{Department of Computer Science, University College London, UK \and
Centre for Artificial Intelligence, University College London, UK\\
\email{\{moiz.imran.22,m.bulathwela\}@ucl.ac.uk}}

\maketitle              
\begin{abstract}
Intelligent tutoring systems increasingly provide automated feedback on student work, but robust feedback requires assessing reasoning, not only final answers. We study a failure mode we call the \emph{correct answer trap (CAT)}: models under-detect misconceptions when students reach a correct answer via flawed reasoning. \textcolor{black}{Analysing real student responses from the Eedi mathematics platform, we show that 71\% of these failures concentrate in just two question types, both sharing a common structure where flawed reasoning happens to produce the correct numerical answer. Comparing a fine-tuned T5 with a frontier large language model, we find that improved capabilities reduce but do not eliminate the problem (84\% vs 57\% detection accuracy). Even the best-performing model generates roughly four false alarms for every genuine detection, making stand-alone screening impractical at realistic class sizes.} Our findings demonstrate that high overall accuracy can mask critical failures in reasoning assessment, and that careful analysis of student reasoning still benefits from human judgment.

\end{abstract}

\section{Introduction}

Intelligent tutoring systems (ITS) promise personalisation at scale, but their effectiveness depends on assessing understanding rather than merely verifying final answers. Students can arrive at the correct answer via flawed reasoning; if a model treats correctness as evidence of understanding, systematic under-detection of misconceptions will prevail. We call this failure mode the \emph{correct answer trap (CAT)}.

For example, a student solving $(-8) - (-5)$ might write: ``8 minus 5 is 3, and since there are negatives, the answer is $-3$.'' The answer is correct, but the reasoning implicitly applies an invalid rule (``subtraction with negatives makes the result negative'') that fails on $(-5) - (-8)$. An answer-checking system would therefore provide false assurance and leave the misconception unaddressed.

This paper characterises the correct answer trap in AI misconception detection. We address three research questions:

\begin{itemize}
    \item \textbf{RQ1 (Where):} In which question types do AI assessors fail most often on correct-answer cases?
    \item \textbf{RQ2 (Why):} \textcolor{black}{Do  specific features make questions vulnerable to this trap?}
    \item \textbf{RQ3 (Models):} Does using stronger reasoning models mitigate this trap?
\end{itemize}

\textcolor{black}{Our contributions are: (1) an empirical characterisation of where CAT cases concentrate in a large student-response dataset collected from a real-world tutoring platform (RQ1), (2) an explanation of why specific questions are vulnerable, grounded in assessment design theory (RQ2), and (3) a model comparison showing that stronger LLMs reduce, but do not close the gap, with analysis of what this means at deployment scale (RQ3). We prioritise depth over breadth, characterising the phenomenon thoroughly within mathematics education.}

\section{Related Work}

\subsubsection{Student Misconception Detection}

\textcolor{black}{Computational approaches to detecting student misconceptions in mathematics are an active area of research. Recent methods include clustering student errors to surface recurring patterns~\cite{gomes2023misconception} and applying NLP to identify misconceptions from student explanations, though with mixed results~\cite{gorgun2023misconception}. Most recent work frames the problem as misconception \emph{classification}: given that a student answered incorrectly, predict which misconception they hold. A large-scale data science competition on the Eedi dataset~\cite{rittlejohnson2025eedi} drew hundreds of teams to this task, and retrieval-augmented approaches have shown promise~\cite{chaudhary2025rag}. Liu et al.~\cite{liu2023novice} evaluate LLMs on the same Eedi dataset, finding that diagnostic accuracy degrades sharply as the number of candidate misconceptions grows. Yet all of these approaches are conditioned on incorrect answers. We address the distinct \emph{detection} problem: determining whether a misconception exists at all, even when the answer is correct. Correct answers remove the signal that classification systems rely upon, and no existing benchmark evaluates this capability.}

\subsubsection{AI Tutoring and Feedback}

AI tutoring systems have shown promise in providing personalised feedback~\cite{vanlehn2011relative}, but most focus on answer correctness rather than reasoning quality~\cite{norris2025next}. Process supervision~\cite{lightman2023lets} demonstrates that reasoning errors occur even when final answers are correct, though this work targets LLM-generated solutions rather than student work. Daheim et al.~\cite{daheim2024stepwise} confirm that stepwise verification of student reasoning remains challenging for frontier models, particularly when the final answer is correct. We build on these findings by identifying \emph{which} questions concentrate failures and what makes them vulnerable.

\subsubsection{Assessment Design and Item Vulnerability}

While AI has been increasingly applied to educational question generation~\cite{li2025novel,bulathwela2023scalable,elkins2024teachers}, far less work has examined what makes questions pedagogically vulnerable. Barton~\cite{barton2018how} identifies a key principle of diagnostic question design: ``it should not be possible to answer the question correctly whilst still holding a key misconception.'' When violated, students applying flawed procedures can coincidentally produce correct answers, a pattern Kapur~\cite{kapur2016examining} terms ``unproductive success.'' Hiebert and Lefevre~\cite{hiebert1986conceptual} distinguish procedural knowledge (executing algorithms) from conceptual knowledge (understanding why procedures work). Procedural questions are more susceptible because rote rules can sometimes produce correct outputs without genuine understanding. We draw on these frameworks to investigate where CAT cases concentrate and why (RQ2).

\section{Methodology}
In this section, we outline the models, datasets and evaluation metrics used in this work.

\subsection{Models}

We compare five model settings belonging to two main categories, (i) smaller models that are easier to finetune, namely,
\textbf{1) T5-small (fine-tuned, encoder-decoder):} A task-specific sequence-to-sequence classifier, fine-tuned for misconception detection,
\textcolor{black}{\textbf{2) BERT-base (fine-tuned, encoder-only):} A standard classification baseline included as a robustness check, confirming the pattern is architecture-agnostic,} and (ii) large foundational models that are \emph{not finetuned} and used as is, namely, 
\textbf{3) Gemini 3 Flash:} A commercial frontier LLM. We evaluate two configurations with low and high thinking budgets.
\textbf{4) Llama-3.3-70B} and
\textbf{5) Llama-3.1-8B:} Open-source LLMs spanning a practical deployment range based on model size.

All prompted models use the same task-specific prompt with explicit guidance about correct-answer cases and use temperature 0 for reproducibility (see supplementary materials)\footnote{Supplementary materials, prompts, and replication scripts found at:   \href{https://github.com/Moiz-I/correct-answer-trap}{github.com/Moiz-I/correct-answer-trap}.}.
We validated this prompt against PedCoT~\cite{jiang2024pedcot} on the 61 TM test cases, finding our task-specific prompt significantly outperforms literature-based prompting (84\% vs 59\% accuracy, McNemar's $p < 0.01$).

\subsection{Data}

\textbf{Primary Dataset:} We use the Eedi\footnote{\href{https://www.eedi.com}{https://www.eedi.com}} benchmark dataset~\cite{rittlejohnson2025eedi}, \textcolor{black}{comprising 20,964 real student responses to mathematics diagnostic questions from a platform used by over 250,000 teachers}. Each response includes the question, correct answer, student answer, and student explanation. Responses were labelled by expert human raters into categories:

\begin{itemize}
    \item \textbf{True\_Correct (TC):} Correct answer with sound reasoning (60\% of data)
    \item \textbf{False-Misconception (FM):} Incorrect answer, misconception evident (38\%)
    \item \textbf{True-Misconception (TM):} Correct answer with flawed reasoning (\textbf{2\%})
\end{itemize}

The TM category ($N = 343$ total; 61 in test set) represents the correct answer trap: rare but pedagogically critical. \textcolor{black}{We evaluate on a stratified 500-sample test set reflecting the natural class distribution. Because this yields only ${\sim}8$ TM cases, we additionally evaluate all 61 TM cases from the same held-out partition. TC and FM metrics come from the 500-sample set; TM metrics come from the full 61-case set.}

\textbf{Question Classification:} Following Hiebert and Lefevre~\cite{hiebert1986conceptual}, we classified each of the 15 unique questions in the Eedi dataset as procedural (n=6), conceptual (n=8), or mixed (n=1) before examining TM distributions over all occurrences of these questions in the dataset. The criterion was whether a memorisable rule could produce correct answers without understanding. Integer subtraction ($(-8)-(-5)$) is procedural: students can apply ``subtract then make negative'' without understanding signed arithmetic. Fraction-of-shape questions are conceptual: students must interpret a visual representation, which no rule can shortcut. Most classifications were unambiguous; only ``dot patterns'' (sequence recognition) warranted the mixed category.

\textbf{Validation Dataset:} \textcolor{black}{To probe generalisability, we also evaluate on PRM800K \cite{lightman2023lets}, a process supervision dataset containing human-labelled reasoning errors in model-generated competition mathematics. We sampled 200 correct-answer cases with reasoning errors and 200 matched wrong-answer cases. Results are reported in the Discussion.}

\subsection{Evaluation Metrics}

\textcolor{black}{We report \textbf{TM recall} (with Wilson 95\% confidence intervals on $n=61$), \textbf{TC/FM recall}, and \textbf{balanced accuracy} (unweighted mean of per-class recalls), which better reflects performance under class imbalance than overall accuracy.}

\section{Results and Discussion}

In this section, we report results relevant to each RQ outlined in the study and draw findings through discussion of results.

\subsection{Distribution of Failures (RQ1)}

Table~\ref{tab:distribution} shows that TM cases concentrate in specific question types rather than distribute uniformly. The top two question types alone account for 71\% of all TM cases.

\begin{table}[b]
\scriptsize
\centering
\caption{Top 5 Question Types by TM Concentration. ``\% of all TM'' shows each type's contribution to the 343 total TM cases.}
\label{tab:distribution}
\begin{tabular}{l l@{\quad}c@{\quad}r@{\quad}r}
\toprule
Rank & Question Type & Class & TM Count & \% of all TM \\
\midrule
1 & Integer subtraction $(-8)-(-5)$ & Procedural & 138 & \textbf{40.2\%} \\
2 & Equivalent fractions $\frac{A}{10} = \frac{9}{15}$ & Procedural & 107 & \textbf{31.2\%} \\
3 & Fraction division $\frac{1}{2} \div 6$ & Procedural & 19 & 5.5\% \\
4 & Dot patterns & Mixed & 17 & 5.0\% \\
5 & Comparing decimals & Conceptual & 16 & 4.7\% \\
\midrule
& \textit{Top 2 combined} & & \textit{245} & \textit{71.4\%} \\
\bottomrule
\end{tabular}
\end{table}

\textcolor{black}{Why these two? In both cases, a common flawed reasoning strategy happens to produce the correct answer for the given numerical values. For $(-8) - (-5)$, students who apply ``8 minus 5 is 3, it's negative so $-3$'' get the right answer despite misunderstanding double negatives. The same rule fails on $(-5) - (-8)$. For $\frac{A}{10} = \frac{9}{15}$, students who spot that $9 - 3 = 6$ or cross-multiply mechanically get $A = 6$ without understanding proportional reasoning. In contrast, $2y = 24$ (also procedural, TM rate 0.06\%) has no such shortcut: misunderstanding equations does not accidentally yield $y = 12$.}

\begin{figure}[t]
\centering
\includegraphics[width=0.7\textwidth]{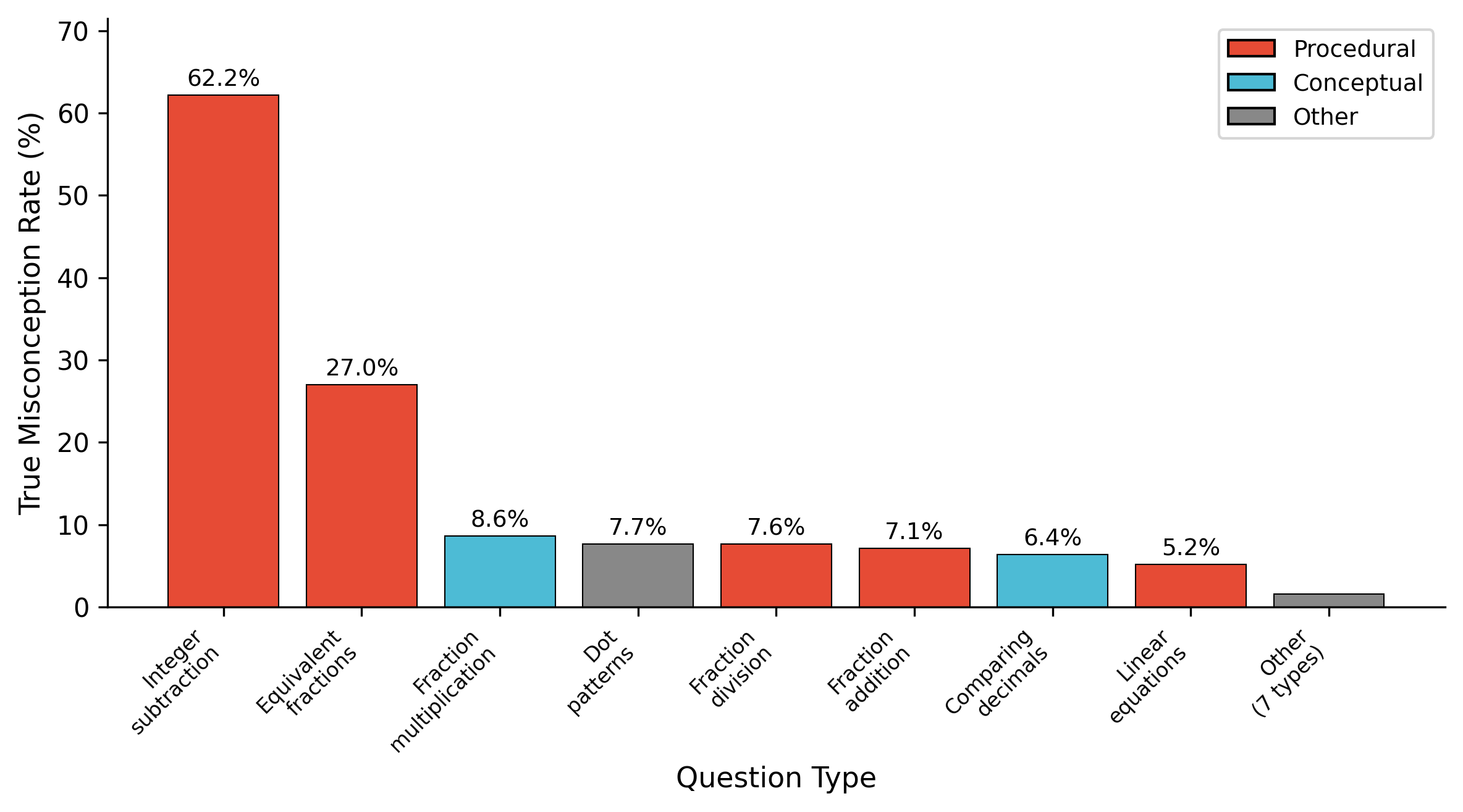}
\caption{TM rate by question type. The top two types (shown in red) account for 71\% of all TM cases.}
\label{fig:distribution}
\end{figure}

\textcolor{black}{\textbf{Finding 1:} CAT cases concentrate in questions where common errors happen to produce the correct answer for the given values.}
Figure~\ref{fig:distribution} visualises this concentration.

\subsection{Item-Level Vulnerability (RQ2)}

\textcolor{black}{Both high-concentration items are procedural, which might suggest that procedural questions are broadly more vulnerable. Removing just these two items, the odds ratio between procedural and conceptual questions drops from 5.6 to 1.0 ($p = 0.48$). The remaining procedural questions have TM rates no different from conceptual ones. The vulnerability belongs to these specific items, not to procedural questions as a class.}

\textcolor{black}{What distinguishes the two vulnerable items is that common flawed procedures happen to produce the correct output for the specific values chosen. Barton~\cite{barton2018how} warns against exactly this in diagnostic question design: if a student can answer correctly while still holding the target misconception, the question fails to diagnose. We term this a \emph{coincidental correctness path}, a form of what Kapur~\cite{kapur2016examining} calls ``unproductive success.''}

\textcolor{black}{\textbf{Finding 2:} CAT concentration is item-driven, not category-driven. The vulnerable items share coincidental correctness paths, identifiable at question-authoring time.}

\subsection{Model Comparison (RQ3)}

Table~\ref{tab:models} compares fine-tuned classifiers (T5 and BERT, trained on the Eedi training split), a frontier LLM (Gemini 3 Flash), and open-source alternatives (Llama 70B and 8B).

\begin{table}[t]
\scriptsize
\centering
\caption{\textcolor{black}{Balanced accuracy (Bal.\ Acc), unweighted mean of per-class recalls along with Per-class recall (\%) for True\_Correct (TC), False\_Misconception (FM), and True\_Misconception (TM) cases. TM evaluated on $n=61$ test cases; Wilson 95\% CIs shown.}}
\label{tab:models}
\begin{tabular}{l@{\quad}r@{\quad}r@{\quad}r@{\quad}r@{\quad}l}
\toprule
Model & \textcolor{black}{Bal.\ Acc} & TC & FM & TM & \textcolor{black}{TM 95\% CI} \\
\midrule
T5-small (fine-tuned) & 83.4\% & 94.9\% & 98.1\% & 57.4\% & [44.9, 69.0] \\
\textcolor{black}{BERT-base (fine-tuned)} & 84.6\% & 99.6\% & 100.0\% & 54.1\% & [41.7, 66.0] \\
Gemini 3 Flash & 87.6\% & 94.0\% & 90.6\% & \textbf{83.6\%} & [72.4, 90.8] \\
Llama-3.3-70B & 63.7\% & 68.3\% & 88.5\% & 81.6\% & [70.0, 89.6] \\
Llama-3.1-8B & 71.0\% & 55.3\% & 74.0\% & 83.6\% & [72.4, 90.8] \\
\bottomrule
\end{tabular}
\end{table}

\textcolor{black}{Fine-tuned models (T5 and BERT) achieve near-perfect separation of correct and incorrect answers (FM recall $\geq$98\%), confirming that the difficulty lies specifically in distinguishing valid from invalid reasoning among correct-answer cases. Both achieve under 58\% TM recall, showing that optimisation for dominant labels leaves rare TM cases under-detected.} Gemini 3 Flash improves TM recall to 83.6\% [72.4, 90.8] while maintaining 94.0\% TC recall; the improvement over T5 is significant (McNemar's test, $p = 0.005$). The Llama models achieve comparable TM recall (81.6--83.6\%) but substantially lower TC recall (55.3--68.3\%), mislabelling sound reasoning as a misconception.
Increasing Gemini's thinking budget slightly reduced balanced accuracy (86.2\% vs 87.6\%), suggesting that longer deliberation can increase over-correction rather than improving detection.

\textcolor{black}{\textbf{Finding 3:} Frontier LLMs achieve the best balance between detecting misconceptions and not over-flagging correct students, but fine-tuned models that can run locally fail on TM. Models that catch more misconceptions also wrongly flag more correct reasoning.}

\subsection{Discussion}

\textcolor{black}{Exploratory integrated gradients analysis suggests that models learn an answer-correctness shortcut~\cite{geirhos2020shortcut}: when the model misclassifies a TM case, answer tokens receive more attribution than explanation tokens (ratio 1.19), while correctly classified TM cases show the reverse (0.68). The model defaults to ``correct answer means correct reasoning.''}

\textcolor{black}{We also tested our best-performing model (Gemini 3 Flash) on PRM800K \cite{lightman2023lets}, a process supervision dataset with human-labelled reasoning errors in model-generated competition mathematics solutions. Detection rates dropped by 10.5 percentage points when final answers were correct (83\% vs 93.5\%, Fisher's exact $p < 0.003$), consistent with our primary findings. However, this dataset contains model-generated solutions rather than student explanations, limiting direct comparison. Whether the correct answer trap extends systematically to other domains and populations remains an open question.}

\subsubsection{Implications for Practice}

Our findings suggest risk-stratified deployment rather than uniform automation. \textcolor{black}{Questions with coincidental correctness paths carry elevated risk: 71\% of TM cases concentrate in just two such items. Question authors can reduce this risk at design time by checking whether known errors happen to yield the correct answer for the chosen values, and revising the values if they do. At natural TM prevalence (1.6\%), even the best model (Gemini 3 Flash) generates roughly 4.3 false alarms per genuine detection, so fully automated screening is not viable on its own. In unstructured classroom settings, where TM prevalence is likely lower still, the false alarm problem would be worse.} Rather than binary classification, follow-up questions that probe understanding may be more practical.
\textcolor{black}{Preliminary experiments suggest that providing mark-scheme-style reference reasoning to frontier models can improve TM detection, similar to how teachers compare student work against expected solution steps. Pairing AI screening with structured references offers a path towards safer deployment.} The study also shows the possibility of developing AI agents that specialise in question vulnerability classification and misconception detection, to serve as micro-skills within an agentic intelligent tutor (e.g. the TrueReason architecture~\cite{truereason}).

\subsubsection{Limitations}
\textcolor{black}{Our task (misconception \emph{detection} on correct answers) is distinct from prior work on the Eedi dataset, which evaluates misconception \emph{classification} on incorrect answers only. Model comparisons use 61 TM samples; at this sample size, Wilson 95\% confidence intervals span $\pm$12 percentage points (Table~\ref{tab:models}). The T5 vs Gemini difference remains significant ($p < 0.005$), but finer distinctions between models should be interpreted cautiously. Population-level findings (RQ1, RQ2) use the full dataset ($N = 343$) and are more robust.}

\textcolor{black}{The question classification (procedural/conceptual) was performed by one researcher following Hiebert and Lefevre 's~\cite {hiebert1986conceptual} criteria over $n=15$ unique questions. RQ2 focuses on item-level concentration rather than category-level prediction, reducing the impact of any classification ambiguity, but we cannot fully separate item identity from question category at this scale. All prompted models used temperature 0; we did not evaluate run-to-run stability across API versions.}

\section{Conclusion}

We characterised the correct answer trap (CAT) in AI misconception detection: the systematic failure to identify flawed reasoning when answers are correct. Our key findings are:

\begin{enumerate}
    \item \textcolor{black}{\textbf{CAT cases are concentrated and predictable.} 71\% occur in just two questions where common student errors happen to yield the correct numerical answer. Removing these two items collapses the procedural/conceptual odds ratio from 5.6 to 1.0. The pattern is about specific items, not broad question categories, and question authors can anticipate it at design time.}
    \item \textcolor{black}{\textbf{Models default to ``correct answer means correct reasoning.''} Fine-tuned models learn near-perfect answer checking (FM recall $\geq$98\%) with almost no confusion between correct and incorrect answers. The only difficulty is distinguishing valid from invalid reasoning when the answer is correct. Exploratory integrated gradients analysis suggests this reflects an answer-correctness shortcut, where answer tokens dominate model predictions for misclassified TM cases.}
    \item \textcolor{black}{\textbf{Better models help, but not enough.} Gemini 3 Flash achieves the best balance (84\% TM recall, 94\% TC recall), but at natural prevalence (1.6\% TM), even this model generates roughly four false alarms per genuine detection. Fine-tuned models that can run locally achieve under 58\% TM recall. Catching these cases currently requires frontier APIs that are too costly for routine classroom use.}
\end{enumerate}

The correct answer trap reflects a broader limitation of answer-focused assessment: high overall accuracy can mask failures precisely where intervention matters most. Preliminary follow-up experiments suggest that moving from binary classification to graduated assessments matching teacher practice can help address this limitation. Future work should expand to other domains and languages and study process supervision approaches that evaluate reasoning steps directly.

\begin{credits}
\subsubsection{\ackname} 
This work is co-funded by the European Commission’s projects “Teacher-AI Complementarity (TaiCo)" (Project ID: 101177268), “Humane AI" (Grant No. 820437) and “X5GON" (Grant
No. 761758).

\end{credits}

%
\bibliographystyle{splncs04}
\bibliography{aied26}

\end{document}